\documentclass[twocolumn,prb,showpacs]{revtex4}
\usepackage{graphicx}

\begin{document}
\title{The disordered flat phase of a crystal surface- critical
 and dynamic properties}
\author{J. Kl\"ars}
\author{W. Selke}
\affiliation{Institut f\"ur Theoretische Physik B,
 RWTH Aachen, 52056 Aachen, Germany}
 
\begin{abstract}
We analyze a restricted SOS model on a square lattice with nearest and
next--nearest neighbor interactions, using Monte Carlo techniques. In 
particular, the critical exponents at the  preroughening transition
between the flat and disordered flat (DOF) phases are confirmed
to be non--universal. Moreover, in the DOF phase, the equilibration
of various profiles imprinted on the crystal
surface is simulated, applying evaporation kinetics and
surface diffusion. Similarities to and deviations from related
findings in the flat and rough phases are discussed.
\end{abstract}


\maketitle

A new type of crystal surface, the disordered flat (DOF) phase, has
been introduced by den Nijs and Rommelse almost twenty years
ago. \cite{RdN,dNR,dN} It is characterized by finite height
fluctuations, although it comprises a disordered array of steps,
thereby combining features of the, conventional, flat and rough
phases. Experimental evidence for a DOF phase has been reported, for
instance, for films of rare gas atoms on graphite and for
GaAs(001). \cite{Youn,Weichman,Ding}

A minimal model to describe the DOF phase has been proposed by
Rommelse and den Nijs. \cite{RdN} It is a restricted SOS, or 
RSOS, model on a square lattice with interactions connecting pairs of
nearest $(i,j)$ and next--nearest $(i,k)$ lattice sites, with the
Hamiltonian 

\begin{equation}
{\cal H} = K\sum\limits_{(i,j)} \delta_{|h_i-h_j|,1} +
  L\sum\limits_{(i,k)} \delta_{|h_i-h_k|,2}
\end{equation}

\noindent 
where $h_i$ is the integer--valued height at site $i$. The absolute
value of the height difference is restricted to be at most one for
neighboring sites, and, thence, at most two for next--nearest
neighbors. $\delta$ is the Kronecker symbol. The interaction
between next--nearest neighbors, $L$, is assumed to be positive.

The phase diagram of the Hamiltonian (1) in the $(k_BT/L,K/L)$--plane is
depicted in Fig.1. It displays five distinct phases: The usual flat
and rough phases, the reconstructed flat and rough phases arising
from a repulsive, negative nearest--neighbor
interaction, $K < 0$, as well as, finally, the DOF phase. The most intriguing
aspects of the phase diagram are the disordered flat phase and the
transition between the flat and the DOF phases, the
preroughening transition. \cite{RdN}

\begin{figure}
  \begin{center}
    \includegraphics[width=0.9\linewidth]{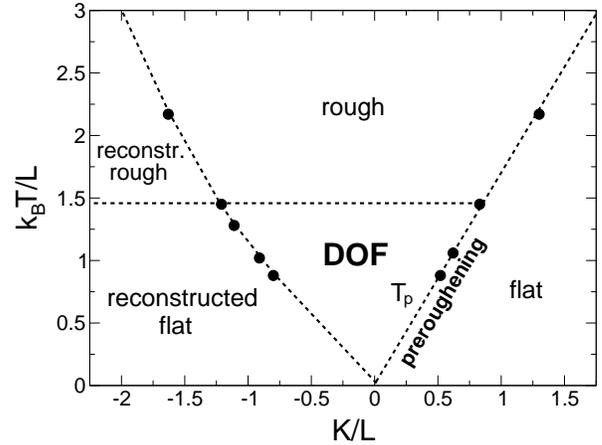}
  \end{center}
  \caption{\label{fig1} Phase diagram of the RSOS model, eq. (1), as
    obtained from a transcription of the diagram determined by
    Rommelse and den Nijs \cite{RdN}, with the circles referring to
    their results using the transfer matrix method. The boundary
    lines of the DOF phase at low temperatures follow from the
    present Monte Carlo simulations.}
\end{figure}

Similar phase diagrams have been obtained for related SOS as well as other
models with short--range interactions \cite{dNR,dN,dN2,Mazzeo},
including variants to describe fcc(110)
surfaces. In the following, we shall first discuss 
critical properties at the preroughening
transition and then dynamic properties of
the DOF phase, analysing the minimal model (1). The findings are
expected to be largely independent of details of the
model. 

To determine critical properties at the preroughening transition, we
used Monte Carlo techniques, in particular, apart from the
standard Metropolis approach, the Wang--Landau and
the transition matrix algorithms. \cite{Wang,Trans} Square systems
of size $M^2$, with $M$ up to 64, have been
simulated. \cite{Klaers} Taking into account finite--size effects, we
estimated standard critical exponents, namely those of the order 
parameter (the parity \cite{dNR}), $\beta$, and its
susceptibility, $\gamma$, as well as 
of the correlation length, $\nu$, at $K/L$= 1/8, 1/4, and
1/2, see Fig. 1.  

Note that the preroughening transition temperature $T_p$, at
fixed ratio $r=K/L$, may be obtained quite accurately by fitting
the height fluctuations to the ansatz $c_{eff} \ln(M)$. Increasing
$M$, $c_{eff}$
tends to zero in the flat and DOF phases, while it seems to approach at
$T_p$ a non-vanishing value, with only minor finite--size
dependences. Thence, the height fluctuations seem to diverge
directly at the preroughening. The prefactor
$c_{eff}(M \longrightarrow \infty, T_p)$ is estimated to be smaller than the
characteristic value $1/\pi^2$ of a transition of
Kosterlitz--Thouless type (which occurs, e.g., when going
directly from the flat to the rough
phase, see Fig. 1).\cite{Klaers}. It is found to increase with
$K/L$, reaching, possibly, $1/\pi ^2$ at the triple point
between the flat, DOF, and rough phases.

The critical exponent $\beta$ is found to tend to increase
with the ratio of the couplings $r= K/L$ from $1.08 \pm 0.09$ at $r$=1/8
to $1.16 \pm 0.02$ at $r$= 1/4, and then to $1.75 \pm 0,03$ at
$r$ =1/2. The corresponding estimates for $\gamma$ are 
$2.75 \pm 0.23$, $2.80 \pm 0.03$, and $3.50 \pm 0.05$. For $\nu$ we
obtain $2.40 \pm 0.20$, $2.52 \pm 0.03$, and $3.43 \pm
0.05$. The rather unusual values of the critical exponents
at the preroughening transition
are in accordance with previous estimates \cite{dNR,Mazzeo}. Our
findings confirm the non--universality of the preroughening
transition \cite{dNR}, with the critical exponents depending
continuously on $K/L$. Applying
scaling laws, the critical exponent
of the specific heat, $\alpha$, turns out to be strongly
negative. Actually, from our simulational data, the specific
heat appears to vary smoothly with temperature, preventing a
direct estimate of $\alpha$.

Dynamic properties of the DOF phase have not been studied in much
detail before, despite a wealth of theoretical and
experimental analyses of morphological changes for flat and rough crystal
surfaces. In fact, we are aware of
only two studies on the equilibration of initially flat surfaces
in the DOF phase, driven by evaporation kinetics, using either Monte
Carlo techniques (studying the Hamiltonian (1) with $K=0$) or
the renormalization group method. \cite{Dev,Park}

In these two studies, the time evolution of the height
fluctuations $<\delta h^2>$ has been
analyzed .\cite{Dev,Park} For systems of $M^2$
sites, one defines
$<\delta h^2> = <\sum\limits_i (h_i - h^*)^2/M^2>$, where
$h^*$ denotes the average height. The height fluctuations
have been observed and argued \cite{Dev,Park} to increase initially
logarithmically in time, both in the rough and DOF phases of
RSOS models. In the DOF phase, then the fluctuations decrease, and 
finally they approach the equilibrium value. In contrast,
in the rough phase, the fluctuations grow monotonically before they 
eventually saturate, for finite $M$.

We confirmed these findings, for $K= 0$ and, in the
DOF phase, $k_BT/L =0.7$ (2.5 in the rough phase), with $M$ ranging
from 16 to 256, using full
periodic boundary conditions (previously \cite{Dev}, helical
boundary conditions had been employed). \cite{Klaers} The
initial height configuration is $h_i(t=0)$=0. To mimic evaporation
kinetics, we apply the standard Metropolis Monte Carlo algorithm,
attempting in each move to change, at a randomly chosen site $i$, the
height by $\pm 1$. In addition to the previous work, we
identified, at very early times, a diffusive behavior,
with $<\delta h^2> \propto t$, due to
uncorrelated height changes at randomly chosen sites. Of course, it
seems doubtful that the logarithmic behavior of
$<\delta h^2>(t)$ will prevail to arbitrarily long times in the
DOF phase when studying larger and larger systems, as it
may be the case in the rough phase. 

The time evolution of the surface may be visualized by
monitoring Monte Carlo configurations. In the DOF phase, at early
times, several clusters of
different local average height, 1/2 or -1/2, are formed, with the
height fluctuations growing logarithmically. In the
clusters, the sites are, in a disordered fashion, predominantly at
heights 0 and 1 or 0 and -1. The subsequent decrease in
$<\delta h^2>$ is due
to the fact that one type of cluster, 1/2 or -1/2, prevails in
the equilibrated DOF phase, see Fig. 2. Of
course, with full periodic boundary conditions, the overall average
height may eventually wander, tending to stick at neighboring
half--integer values. This effect, with the entire surface undergoing a 
random walk, does not affect the height fluctuations.     

\begin{figure}
  \begin{center}
    \includegraphics[width=0.8\linewidth]{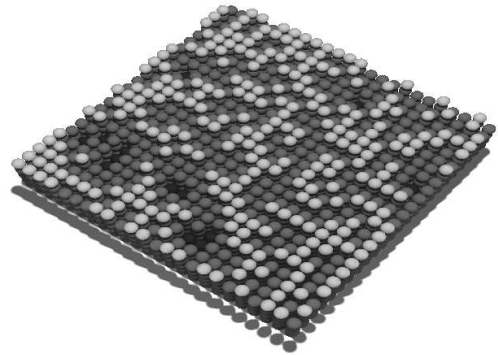}
  \end{center}
  \caption{\label{fig2} Typical equilibrium Monte Carlo configuration
    in the DOF phase, with $K=0$ and $k_BT/L= 0.1$, comprising sites
    at height -1 (black), 0 (dark grey), and 1 (light grey). The surface
    has $24^2$ sites.}
\end{figure}

To analyze more completely the dynamics in the DOF phase, we
also studied the equilibration
of various other one--dimensional profiles 
imprinted on the surface, in particular, steps, periodic gratings, and
wires. \cite{Klaers} There, the interplay of the orientation
dependent surface tension and mobility \cite{Mullins,Spohn} leads
to interesting
morphological changes in the rough and flat phases, for evaporation
kinetics as well as for surface
diffusion, see pertinent reviews on theory and
experiments. \cite{Pimpinelli,Williams,Giesen,Bonzel} 

In the DOF phase, a step of monoatomic height may be introduced by
fixing the average heights of the sites at two opposite boundaries to
be, e.g., at 1/2 or 3/2, by setting the heights at these boundary sites 
alternately 0 and 1, or 1 and 2, respectively. The other two
boundaries are connected by periodic boundary conditions. Square
systems with $M^2$ sites
are considered. Averaging over sites parallel to the
fixed boundaries as well as over many, $N$, realizations of
the equilibration process, one obtains a one--dimensional height
or step profile $H(x,t)$, with $x$ running
from, say, $-M/2+1/2$ to $M/2-1/2$. At time $t=0$, the
height configuration is assumed to look like
a checkerboard comprising
heights 0 and 1 in the half of the surface next to the 1/2--boundary,
and heights 1 and 2 in the other half. Then, $H(x,t=0)= 1/2$
for $x < 0$, and $H(x,t=0)$= 3/2 for $x > 0$, with a sharp
step in between. As time 
proceeds, the step profile will start to broaden, see
Fig. 3. In the flat and rough phases, the step width is known \cite{Giesen}
to grow asymptotically in
time, $t$ and in the thermodynamic limit, $M \longrightarrow \infty$,
with  a power law $\propto t^b$. The characteristic exponent
$b$ depends on the transport mechanism and
the type of phase. For instance, in the case of
evaporation kinetics, $b$ is 1/2 in the rough and
1/4 in the flat phase, in the case of surface diffusion, $b$ is 1/4
in the rough and 1/6 in the
flat phase.\cite{SelDux,Einstein,Blago,SeSzHa,Upton}. $b$ may be
determined from Monte Carlo 
data by monitoring the position of the step profile
at fixed height. \cite{SeSzHa} In that way, an
effective exponent $b_{eff}$ may be estimated, approaching the true
value $b$ in the limits $M,t \longrightarrow \infty$.

\begin{figure}
  \begin{center}
    \includegraphics[width=0.80\linewidth]{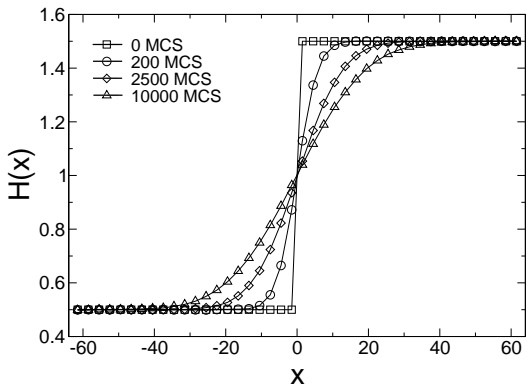}
  \end{center}
  \caption{\label{fig3} Evolution of step profiles, $H(x,t)$, in
    the DOF phase, $K/L=0$, $k_BT/L= 0.1$, at various times (measured in
    Monte Carlo steps per site, MCS), applying evaporation
    kinetics. Systems of $128^2$ sites have been
    simulated, averaging over $N= 11200$ realizations. For clarity, only
    each third profile point, on both sides of the step, is shown. }
\end{figure}

Typical simulated step profiles, applying evaporation kinetics, in
the DOF phase are depicted in Fig. 3. Analysing such profiles, we
obtain an effective exponent, for $M= 128$, approaching, from 
above, closely 1/2 in the rough phase. In the DOF
phase ($M= 128, K=0, k_BT/L= 0.1$), the effective exponent also 
decreases monotonically with time, the data being
well compatible with an asymptotic value $b= 1/4$, as in the
flat phase. Surface diffusion may be mimicked in   
the simulations by attempting
to, say, decrease the height at a randomly chosen site by one 
and to increase
simultaneously the height at a neighboring site by one. For that
dynamics, the effective $b$ is found to
approximate nicely 1/4 in the rough phase, $M= 128$, and to approach
1/6 in the DOF phase, $M=256, K=0, k_BT/L= 0.1$.

An interesting phenomenon occurs when periodic, one--dimensional
gratings are imprinted on the surface. In the flat phase, the
healing proceeds layer by layer, driven by an annihilation of
the meandering steps bordering the top and bottom
terraces of the
grating. \cite{SelBie,SelDux,Ebner,Erle,Kandel,Karim}. To study
the equilibration of gratings, especially in
the DOF phase, we consider an initial profile of the form  
$H(x,t=0)= aint(A \sin(2\pi x/M))$, i.e., a
discretized sine function with amplitude A. Full periodic
boundary conditions are employed for lattices of $M^2$ sites. We
first shall present results on evaporation kinetics, with
wavelength $M$ ranging from 32 to 512.

\begin{figure}
  \begin{center}
    \includegraphics[width=1.0\linewidth]{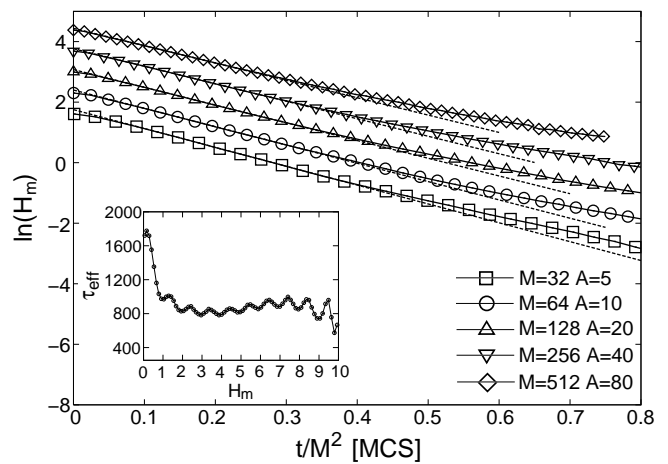}
  \end{center}
  \caption{\label{fig4} Time, measured in Monte
    Carlo steps per site, divided by $M^2$, dependence 
    of the amplitude $H_m$ of
    sinusoidal gratings for various system sizes, $M$, and initial
    amplitudes $A$, setting $K=0$ and $k_BT/L$= 0.7. Evaporation
    kinetics is simulated, averaging over several (ranging from $N=10$
    for $M= 512$ up to $N=48000$ for $M=32$) realizations. The dashed lines
    correspond to an exponential decay of the amplitude $H_m$. In the
    inset, the effective relaxation time, $\tau_{eff}$, as a function
    of $H_m$ is shown for $M=64, A=10$. }
\end{figure}

In the DOF phase, $K=0$ and, mostly, $k_BT/L= 0.7$, the profile of the
gratings is quite close to sinusoidal during the
equilibration, showing, however, in the beginning a tendency towards a
broadening near the extrema. The time
dependence of the decay of the
amplitude, $H_m(t)$, is illustrated in Fig. 4, studying profiles
with fixed ratio $M/A$. From this figure and
other simulational data, we infer that the amplitude
falls off at fairly early times approximately
exponentially, $H_m \propto \exp(-t/\tau)$, $\tau$ being the
relaxation time, in accordance with the prediction
of Mullins \cite{Mullins} for the rough phase. In that time
regime, $\tau$ is proportional to $M^2$, also as
predicted. \cite{Mullins} However, as time goes on, in contrast to the
behavior in the rough phase,  a systematic
deviation from the exponential decay occurs, with a 
significantly slower equilibration. When fitting there the decay to 
a power--law, the effective exponent depends both on time and
the wavelength $M$ (fixing $M/A$), becoming lower for larger
systems (in the largest system we studied, $M=512$, we find an
effective exponent of roughly $-1.5$ to $-2$). Note that one might be still
rather far from a characteristic asymptotic temporal dependence. For
comparison, in the
flat phase, the equilibration is predicted \cite{Lancon} to follow, at long
times, $H_m \propto t^{-1/2}$. The deviation 
from the exponential form sets in at a larger
amplitude for gratings with larger wavelength and larger initial
amplitude $A$, see Fig. 4. Indeed, that amplitude seems
to grow like $M^{\delta}$,
with $\delta$ being about 5/4. Thence, we tend to conclude, that the
theory of Mullins describes merely a transient behavior
for gratings of finite wavelength in the DOF phase.

By inspection of Monte Carlo configurations during the
relaxation, one observes the decay to proceed layer by
layer. At a reference moment of time, the top (and bottom) terrace
of the grating may display a flat disordered structure, with the sites  
being predominantly at the integer heights $n$ and $n-1$, yielding an 
amplitude $H_m$ of approximately $n-1/2$. The terrace is now  bordered
by two meandering steps, and the amplitude of the profile (as
it follows from averaging over many realizations) may
decrease fairly slowly, when time goes on. Eventually the
two steps may touch
each other, giving rise to islands of flat disordered structures
on top of another flat disordered terrace of average height
$n-3/2$. The islands then shrink and dissolve quite
quickly, associated with a rather fast
decay of the amplitude. Thence, we expect two time scales, due
to the meandering of the steps and the dissolving of islands, in
accordance with previous simulational results on the
equilibration of gratings in the
flat phase. \cite{SelBie,Ebner}. Indeed, as exemplified in the inset
of Fig.4, the two time scales lead to an oscillatory behavior in the 
effective relaxation time, $\tau _{eff} = -(d\ln H_m/dt)^{-1}$, as
a function of the amplitude $H_m$. The
pronounced increase of $\tau _{eff}$ at small
amplitudes corresponds to the onset of the 
slower, non--exponential decay, discussed
above. In the rough phase, these oscillations tend to fade away, for
obvious reasons.

In the case of surface diffusion, similar features hold. We set
$K=0$ and $k_BT/L$= 0.7 in the DOF phase ($k_BT/L= 2.5$ in the
rough phase), with $A$ ranging from
3 to 6, and with $M$ ranging from  24 to 48. The equilibration
proceeds on much longer time scales than in the evaporation
case. First, as it is true for evaporation kinetics as well, the discretized
initial configuration relaxes to a close--to--sinusoidal profile with 
disordered flat structures at the extrema. Then, still at
early times, the decay is
described well by an exponential time dependence, with the relaxation
time being proportional to $M^4$, in agreement with Mullins' theory
for surface diffusion in the rough phase. \cite{Mullins} Eventually,
as time goes on, in the DOF phase a slowing down sets in, like for
evaporation--recondensation. A crossover
from an exponential decay to a slower decay at larger times had
been found before for surface diffusion in the flat
phase, with the crossover time depending on temperature. \cite{Ebner}

Finally, we studied the flattening of one--dimensional wire
profiles. In the DOF phase, applying evaporation kinetics, the
equilibration proceeds again layer by layer, driven by step annihilation at
the disordered flat top terraces, accompanied by two time
scales. Analogous features have been found before for the
flat phase \cite{SelDux}. The excess mass
at the surface, due to the wire, changes with time, with
the average height tending to approach a half--integer value. In
contrast, in the rough phase, the excess
mass is nearly conserved. In the rough phase, the amplitude of
the wire approaches a decay $H_m \propto t^{-1/2}$, as 
predicted by the Mullins' theory and already confirmed in
previous simulations. To determine the asymptotic decay
law in the flat \cite{SelDux} and DOF phases, presumably wires with
very large initial width and amplitude are needed, being beyond the
reach and scope of the present study.

In summary, the preroughening transition is found to be
non--universal. Dynamic properties in the DOF
phase resemble partly ones observed
in the flat phase, partly ones observed in the rough phase. They
deserve to be studied in even more detail in the future.   
  
 


\end{document}